\documentclass[conference, 10pt]{IEEEtran}
\IEEEoverridecommandlockouts
\usepackage{cite}
\usepackage{amsmath,amssymb,amsfonts}
\usepackage{algorithmic}
\usepackage{graphicx}
\usepackage{textcomp}
\usepackage{xcolor}
\usepackage{booktabs}
\ifCLASSOPTIONcompsoc
    \usepackage[caption=false, font=normalsize, labelfont=sf, textfont=sf]{subfig}
\else
    \usepackage[caption=false, font=footnotesize]{subfig}
\fi
\usepackage{pgfplots}%
\pgfplotsset{compat=1.13}
\usepackage{circuitikz}
\usepackage{tikz-timing}
\usetikztiminglibrary{overlays}
\usepgflibrary{shapes.gates.flipflops.IEC}

\definecolor{fgblue}{RGB}{145,191,219}%
\definecolor{fgred}{rgb}{0.9,0,0}%
\definecolor{mybarcolor}{RGB}{145,191,219}

\begin{document}

\onecolumn

\noindent \textcopyright{} 2019 IEEE. Personal use of this material is permitted. Permission from IEEE must be obtained for all
other uses, in any current or future media, including reprinting/republishing this material for advertising or
promotional purposes, creating new collective works, for resale or redistribution to servers or lists, or reuse
of any copyrighted component of this work in other works.

\twocolumn

\title{Functional Failure Rate Due to Single-Event Transients in Clock Distribution Networks\\
\thanks{This work was supported by the RESCUE project which has received funding from the European Union's Horizon 2020 research and innovation programme under the Marie Sklodowska-Curie grant agreement No. 722325.}
}

\author{%
\IEEEauthorblockN{%
  Thomas Lange\IEEEauthorrefmark{1}\IEEEauthorrefmark{2},
  Maximilien Glorieux\IEEEauthorrefmark{1},
  Dan Alexandrescu\IEEEauthorrefmark{1},
  Luca Sterpone\IEEEauthorrefmark{2}%
}
\IEEEauthorblockA{%
  \IEEEauthorrefmark{1}\textit{iRoC Technologies}, Grenoble, France \\
  \IEEEauthorrefmark{2}\textit{Dipartimento di Informatica e Automatica, Politecnico di Torino}, Torino, Italy \\
  \{thomas.lange, maximilien.glorieux, dan.alexandrescu\}@iroctech.com \qquad
  luca.sterpone@polito.it}
}

\maketitle

\begin{abstract}
With technology scaling, lower supply voltages, and higher operating frequencies clock distribution networks become more and more vulnerable to transients faults. These faults can cause circuit-wide effects and thus, significantly contribute to the functional failure rate of the circuit. This paper proposes a methodology to analyse how the functional behaviour is affected by Single-Event Transients in the clock distribution network. The approach is based on logic-level simulation and thus, only uses the register-transfer level description of a design. Therefore, a fault model is proposed which implements the main effects due to radiation-induced transients in the clock network. This fault model enables the computation of the functional failure rate caused by Single-Event Transients for each individual clock buffer, as well as the complete network. Further, it allows the identification of the most vulnerable flip-flops related to Single-Event Transients in the clock network.

The proposed methodology is applied in a practical example and a fault injection campaign is performed. In order to evaluate the impact of Single-Event Transients in clock distribution networks, the obtained functional failure rate is compared to the error rate caused by Single-Event Upsets in the sequential logic.
\end{abstract}


\section{Introduction}

Today's reliability standards and customers' expectations set tough
targets for the quality of electronic devices and systems. Among other
reliability threats, transient faults, such as Single-Event Upsets (SEUs) in
sequential/state logic and Single-Event Transients (SETs) in combinatorial
logic, are known to contribute significantly to the overall failure
rate of the system, possibly exceeding the set reliability targets. As
an example, standard flip-flops and SRAM memories, manufactured in
relatively recent technologies (down to the latest CMOS bulk
processes) exhibit error rates of hundreds of FITs (events per a
billion working hours per
megabit)~\cite{baumann_radiation-induced_2005, seifert_radiation-induced_2006}. Complex
circuits using such cells can easily overshoot the by ISO~26262
mandated 10\,FIT target for an automotive ASIL~D application.

Circuits' susceptibility to transient faults/single events is caused
by faults occurring in the circuit's cells and their subsequent
propagation in the system, possibly causing observable effects
(failures) at the system level. The impact of Single-Event Upsets and
Single-Event Transients in individual state and combinatorial cells
has been extensively studied and
for many applications, is the leading contributor to the overall event
rate exhibited by the circuit. However, due to technology scaling,
lower supply voltages and higher operating frequencies, other circuit
features such as the clock distribution network (CDN), reset circuitry,
etc. become also more vulnerable to transient
faults~\cite{dodd_basic_2003, shivakumar_modeling_2002, wissel_flip-flop_2009, chellappa_90-nm_2012} and
could cause circuit-wide effects that are more difficult to mitigate
and to correct. Indeed, clock buffers from the clock distribution networks
have a high fan-out and very few masking mechanism; Single-Event
Transients occurring in these cells can potentially reach many
sequential cells and state elements and thus, significantly contribute
to the overall functional failure rate.

\subsection{Objective of Our Methodology}

So far, only few works studied the impact of SETs in clock networks. To determine the sensitivity of clock buffer cells to these events, some studies performed accelerated radiation tests of dedicated test chips~\cite{wissel_flip-flop_2009, malherbe_investigating_2016}. Other approaches computed a static failure rate by performing circuit simulation on the electrical-level and thus, obtaining the Electrical De-Rating per clock buffer, as well as the upset rate of the sequential logic due to SETs in the clock network. This upset rate was combined with the functional failure rate due to SEUs in the sequential logic obtained from a SEU fault injection campaign~\cite{chipana_soft-error_2012, chipana_set_2012}. However, their SET fault injection simulations used only static inputs and thus do not reflect any dynamic behaviour during the runtime of the circuit. Hence, \cite{hao_single-event_2017} extended this method by injecting SETs in the clock distribution network during a dynamic electrical simulation and thus, obtaining the faulty latching activity of the sequential logic.

Nonetheless, the previous work does not analyse the impact of SETs on the functional behaviour of the circuit and furthermore, they are all based on electrical simulations. Since the complexity of today's circuits is increasing, a dynamic simulation of the full circuit on the electrical level is not feasible anymore. Thus, contrary to the previous work, the proposed fault model in this paper is based on logic-level simulation and only requires the register transfer level description of a design. This enables a faster analysis of the circuit. The proposed method is evaluated by applying it on a practical example and performing a fault injection campaign.

\subsection{Organisation of the Paper}

The remainder of this paper is organised as follows: Section~\ref{sec:background} summarises the definition of Single-Event Effects and the different de-rating mechanism and relates them to the context of SETs in the clock distribution networks. The proposed methodology and dedicated fault model are described in section~\ref{sec:methodology}. In section~\ref{sec:results} the proposed method is validated on a practical example and the functional failure rate for each clock buffer and the whole network are computed. Further, the most vulnerable flip-flops related to transients in the clock network are identified. Section~\ref{sec:conclusion} rounds off this paper by giving concluding remarks as well as prospects for future work.
\section{Single-Event Effect Mechanism with Regard to Clock
  Distribution Networks}
\label{sec:background}

Erroneous data in one of the memory or logic points of
a circuit can be produced by the propagation of a Single-Event
Transient (SET) or Single-Event Upset (SEU). SETs are the result
of the collection of charge deposited by ionising particles on
combinatorial logic cells. SEUs are the change of the logic state of a
discrete sequential element, such as a latch, a flip flop or a memory
cell.

In the data path between flip-flops, four de-rating
mechanisms~\cite{nguyen_systematic_2003, alexandrescu_towards_2012} significantly reduce the impact of SETs and SEUs on the effective error rate.
\begin{LaTeXdescription}
\item [Electrical De-Rating (EDR):] The transient is filtered due to pulse
  narrowing and or an increase of the rise and fall time during its
  propagation. By the time it reaches the end of the path, either it
  has been completely filtered or the voltage transition is below the
  switching threshold.
\item [Temporal De-Rating (TDR):] The erroneous state reaches the
  input of a flip-flop but outside the latching window, thus it is not sampled.
\item [Logical De-Rating (LDR):] The erroneous state is prevented from
  propagating due to the state on another controlling input of a gate
  such as a zero value on an \verb+AND2+ gate.
\item [Functional De-Rating (FDR):] The erroneous state is considered
  at an applicative level. This means even when an SEU/SET does
  propagate (e.g. is not logically or temporally masked), the impact
  at the function of the circuit can vary, and in many cases is
  benign. Thus, considering the faults at an applicative level, the
  de-rating depends on the criteria defining the acceptable behaviour
  of the circuit during the execution of an application and the fault
  classifications (correctable, uncorrectable, not detected by the
  hardware but detected by the software, if a retry is possible, if
  there is a time limit to receive the correct result, etc.)
\end{LaTeXdescription}
These structural de-ratings mechanism are used to evaluate the probability
of the propagation of a fault during the clock cycle of their
occurrence. They are usually estimated by using probabilistic algorithms and
simulation based approaches.

For a transient in a clock distribution network (CDN), the Logical De-Rating
and Temporal De-Rating is limited. Potentially, an SET may be logically
masked by a clock gating cells or an enable pin of a
flip-flop. Temporal De-Rating is limited as the clock input of the
flip-flop is by definition asynchronous.

In~\cite{seifert_radiation-induced_2005} two main effects are
identified due to transients in the clock network: radiation-induced
jitter and radiation-induced race. Jitter occurs if an transient
causes the clock edge to move forward or backward causing a timing violation. A
race condition occurs if a transient causes a flip-flop that is closed to
become open allowing data to ``race'' through to the next stage.

The objective of this paper is to present a methodology to compute the functional failure rate of a circuit with regards to Single-Event Transients in the clock network. Therefore, the described radiation-induced effects are implemented in a fault model based on logic-level simulation which is presented in the next section.
\section{Methodology}
\label{sec:methodology}

To analyse how the functional behaviour is affected by SETs in the clock distribution network (CDN), the main radiation-induced effects are implemented in a fault model. In order to cope with the complexity of today's circuits the proposed fault model is based on logic-level simulation, which enables a faster analysis than simulations based on the electrical level. By using this fault model in a fault simulation campaign the functional failure rate for each clock buffer and the whole network can be calculated. Further, the vulnerability of the sequential logic in relation to these events can be computed.

\subsection{Fault Model}

\begin{figure}[htbp]
    \centering
    
    \subfloat[%
        Procedure to inject an SET into a clock buffer%
        \label{fig:virt_clk_tree}%
    ]{%
        \resizebox{\linewidth}{!}{\definecolor{myblue}{RGB}{145,191,219}
\definecolor{myred}{RGB}{252,141,89}
\definecolor{mygreen}{RGB}{255,255,191}
\definecolor{mygray}{RGB}{210,214,221}

\begin{circuitikz}[
    dff-clk/.style={
        dff, 
        flip flop bottom clock,
    },
]
    \node [
        draw, thick,
        rectangle, scale=3,
        label=below:\shortstack{Clock \\ Source}
    ] (clk-src) 
        at (0,0) {};
        
    \node [buffer, thick] (buffer1) 
        at ($(clk-src) + (2,0)$) {};
    
    \node [buffer, thick] (buffer1-1) 
        at ($(buffer1) + (3,+5.5)$) {};
    \node [buffer, thick] (buffer1-2) 
        at ($(buffer1) + (3,-5.5)$) {};
    
    \pgfmathsetmacro{\bufx}{3}
    \node [buffer, thick] (buffer1-1-1) 
        at ($(buffer1-1) + (\bufx,+2.25)$) {};
    \node [buffer, thick] (buffer1-1-2) 
        at ($(buffer1-1) + (\bufx,-2.25)$) {};
    
    \node [buffer, thick] (buffer1-2-1) 
        at ($(buffer1-2) + (\bufx,+2.75)$) {};
    \node [buffer, thick] (buffer1-2-2) 
        at ($(buffer1-2) + (\bufx,-2.75)$) {};
    
    \pgfmathsetmacro{\ffx}{3.75}
    \node [dff-clk] (dff1-1-1-1) 
            at ($(buffer1-1-1) + (\ffx,+1.05)$) {};
    \node [dff-clk] (dff1-1-1-2) 
            at ($(buffer1-1-1) + (\ffx,-1.05)$) {};
    
    \node [dff-clk] (dff1-1-2-1) 
            at ($(buffer1-1-2) + (\ffx,+1.05)$) {};
    \node [dff-clk] (dff1-1-2-2) 
            at ($(buffer1-1-2) + (\ffx,-1.05)$) {};
    
    \node [dff-clk] (dff1-2-1-1) 
            at ($(buffer1-2-1) + (\ffx,+2.1)$) {};
    \node [dff-clk] (dff1-2-1-2) 
            at ($(buffer1-2-1) + (\ffx,+0)$) {};
    \node [dff-clk] (dff1-2-1-3) 
            at ($(buffer1-2-1) + (\ffx,-2.1)$) {};
    
    \node [dff-clk] (dff1-2-2-1) 
            at ($(buffer1-2-2) + (\ffx,+1.05)$) {};
    \node [dff-clk] (dff1-2-2-2) 
            at ($(buffer1-2-2) + (\ffx,-1.05)$) {};
    
    \foreach \x in {1,2}{
        \foreach \y in {1,2}{
            \foreach \z in {1,2}{
                \draw[thick] (dff1-\z-\y-\x.D) -- +(-0.5,0)
                    node[left] {$D_\text{in}$};
                \draw[thick] (dff1-\z-\y-\x.Q) -- +(0.5,0)
                    node[right] {$Q_\text{out}$};
            }
        }
    }
    \draw[thick] (dff1-2-1-3.D) -- +(-0.5,0)
        node[left] {$D_\text{in}$};
    \draw[thick] (dff1-2-1-3.Q) -- +(0.5,0)
        node[right] {$Q_\text{out}$};

    \draw[thick] (clk-src) -- (buffer1.in);
    
    \draw[thick] (buffer1.out) -- +(0.75,0) |- (buffer1-1.in);
    \draw[thick] (buffer1.out) -- +(0.75,0) |- (buffer1-2.in);
            
    \pgfmathsetmacro{\bufoutx}{1.25}
    \draw[thick] (buffer1-1.out) -- +(\bufoutx,0) |- (buffer1-1-1.in);
    \draw[thick] (buffer1-1.out) -- +(\bufoutx,0) |- (buffer1-1-2.in);
    \draw[thick] (buffer1-2.out) -- +(\bufoutx,0) |- (buffer1-2-1.in);
    \draw[thick] (buffer1-2.out) -- +(\bufoutx,0) |- (buffer1-2-2.in);
        
    \pgfmathsetmacro{\bufoutx}{1}
    \draw[thick] (buffer1-1-1.out) -- +(\bufoutx,0) |- (dff1-1-1-1.Clk);
    \draw[thick] (buffer1-1-1.out) -- +(\bufoutx,0) |- (dff1-1-1-2.Clk);
    \draw[thick] (buffer1-1-2.out) -- +(\bufoutx,0) |- (dff1-1-2-1.Clk);
    \draw[thick] (buffer1-1-2.out) -- +(\bufoutx,0) |- (dff1-1-2-2.Clk);
    \draw[thick] (buffer1-2-1.out) -- +(\bufoutx,0) |- (dff1-2-1-1.Clk);
    \draw[thick] (buffer1-2-1.out) -- +(\bufoutx,0) |- (dff1-2-1-2.Clk);
    \draw[thick] (buffer1-2-1.out) -- +(\bufoutx,0) |- (dff1-2-1-3.Clk);
    \draw[thick] (buffer1-2-2.out) -- +(\bufoutx,0) |- (dff1-2-2-1.Clk);
    \draw[thick] (buffer1-2-2.out) -- +(\bufoutx,0) |- (dff1-2-2-2.Clk);
    
    \coordinate (stage1-sep) at ($(buffer1) !0.5! (buffer1-1)$);
    \path (stage1-sep) -- 
        (stage1-sep |- dff1-1-1-1.north) -- 
        +(-0.5,1.5) coordinate (stage1-sep-top);
    \draw[very thick, dashed, mygray] 
        (stage1-sep-top) -- 
        ($(stage1-sep-top |- dff1-2-2-2.south) - (0,0.25)$);
    \coordinate (stage2-sep) at ($(buffer1-1) !0.5! (buffer1-1-1)$);
    \path (stage2-sep) -- 
        (stage2-sep |- stage1-sep-top) -- 
        +(0,0) coordinate (stage2-sep-top);
    \draw[very thick, dashed, mygray] 
        (stage2-sep-top) -- 
        ($(stage2-sep-top |- dff1-2-2-2.south) - (0,0.25)$);
    \coordinate (stage3-sep) at ($(buffer1-1-1) !0.5! (dff1-1-1-1)$);
    \path (stage3-sep) -- 
        (stage3-sep |- stage1-sep-top) -- 
        +(-0.5,0) coordinate (stage3-sep-top);
    \draw[very thick, dashed, mygray] 
        (stage3-sep-top) -- 
        ($(stage3-sep-top |- dff1-2-2-2.south) - (0,0.25)$);
    
    \node[align=center] (stage2-text) 
        at ($(stage1-sep-top) !0.5! (stage2-sep-top) - (0,0.5)$) 
        {Buffer Stage 2};
    \node[align=center] (stage3-text) 
        at ($(stage2-sep-top) !0.5! (stage3-sep-top) - (0,0.5)$)
        {Buffer Stage 3};
    \coordinate (stage1-mid) at ($(clk-src) !0.5! (buffer1)$);
    \node[align=center] (stage1-text) 
        at (stage1-mid |- stage2-text)
        {Buffer Stage 1 / \\ Root};
    \node[align=center] (ffs-text) 
        at (dff1-1-1-1 |- stage2-text)
        {End-Point / Flip-Flop};
    
    \node [
        draw, rectangle, thick, fill=myblue,
        minimum width=5.25cm, inner sep=10pt
    ] (select-buf) at ($(clk-src) + (-1,9)$) 
        {Select Clock Buffer};
    \node [
        draw, rectangle, thick, fill=mygreen,
        minimum width=5.25cm, inner sep=10pt
    ] (identify-ffs) at ($(select-buf) - (0,1.75)$) 
        {Identify Connected FFs};
    \node [
        draw, rectangle, thick,
        minimum width=5.25cm, inner sep=7.5pt, align=center
    ] (find-sig-names) at ($(identify-ffs) - (0,2)$) 
        {Find Corresponding \\ Signal Names in RTL Model};
    \node [
        draw, rectangle, thick, fill=myred, 
        minimum width=5.25cm, inner sep=7.5pt, align=center
    ] (copy-values) 
        at ($(find-sig-names) - (0,2)$) 
        {Copy $D_\text{in}$ Signal Value \\ to $Q_\text{out}$ Output Signal};
    
    \draw[->, >=stealth, thick] (select-buf) -- (identify-ffs);
    \draw[->, >=stealth, thick] (identify-ffs) -- (find-sig-names);
    \draw[->, >=stealth, thick] (find-sig-names) -- (copy-values);
    
    \node [
        draw, rectangle, scale=7, thick,
        dashed, fill=myblue, opacity=0.5
    ] (select-buf-box) at (buffer1-1) {};
    
    \filldraw[
        draw, dashed, thick,
        fill=mygreen, opacity=0.5
    ]   (buffer1-1.out) -- 
        ($(dff1-1-1-1.north) + (0,0.5)$) --
        ($(dff1-1-2-2.south) - (0,0.5)$) -- cycle;
    
    \draw[
        draw, dashed, thick,
        fill=myred, opacity=0.5
    ] (buffer1-1 -| dff1-1-1-1) ellipse (1.5 and 5);
    
\end{circuitikz}}
    }
    
    \subfloat[%
        Affected flip-flop (state changed)%
        \label{fig:set_timing_ff_change}%
    ]{%
            \input{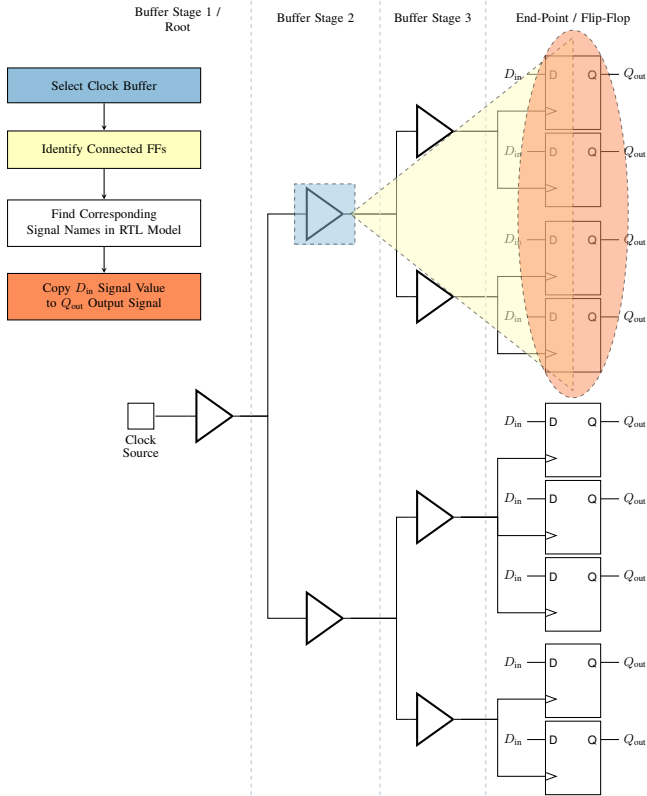}
    }
    
    \subfloat[%
        Unaffected flip-flop (state unchanged)%
        \label{fig:set_timing_ff_unchange}%
    ]{%
            \input{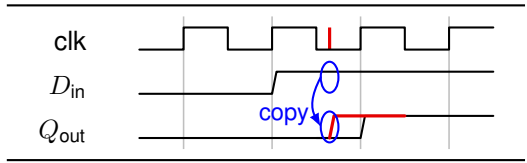}
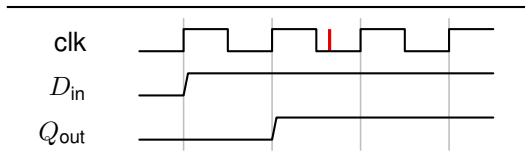
    }
    \caption{Proposed fault model for Single-Event Transients in clock distribution networks based on logic level simulation}
    \label{fig:fault_model}
\end{figure}

The proposed fault model which implements effects of Single-Event Transients propagating along the clock network is illustrated in Fig.~\ref{fig:virt_clk_tree}. It is based on logic-level simulations and thus, only uses the register-transfer level (RTL) description of a design. To emulate the SET in the clock network, first, a clock buffer is selected as injection target. Second, all flip-flops which are connected to the end-point of the selected clock buffer are identified. Then, during the RTL simulation, for each identified flip-flop, the corresponding signal values at the flip-flop output are modified at the injection time. The SET induced clock pulse is imitated by copying the signal value from the flip-flop input signal $D_\text{in}$ to its output signal $Q_\text{out}$ as shown in Fig.~\ref{fig:set_timing_ff_change}. Thus, only flip-flops which would have changed their state in the following clock cycle are impacted by the transient and others remain unchanged (as shown in Fig.~\ref{fig:set_timing_ff_unchange}). 

The proposed fault model does not take any electrical or timing behaviour into account and thus, is representing a worst case scenario. However, it can be combined with measured cross-sections of the clock buffer cells obtained during radiation experiments, as shown in~\cite{malherbe_investigating_2016}, or Electrical De-Rating factors obtained from electrical level simulations (without taking the runtime behaviour into account) as described in~\cite{chipana_soft-error_2012}.

\subsection{Virtual Clock Network}

The proposed method relies on the RTL model of a design. Typically, These models do not provide a clock distribution network. In general, the clock network is obtained by performing a clock network synthesis during the physical design stage of a chip. In this paper this step is simplified by generating a virtual clock network. The generation of a virtual clock network enables an analysis of the circuit in earlier design stages with regard to clock network issues (such as SETs) and allows the evaluation of different clock network features, such as the fan-out, layout or topology. In fact, recent work has shown that topology and gate load play a significant role in the overall SET sensitivity of clock networks~\cite{wang_single-event_2016}.

In the most common implementation of clock distribution networks buffers are inserted either at the clock source and/or along the clock path, forming a tree structure. Thereby, the most used topology of the networks is the symmetric H-tree which can be also seen as a binary tree topology~\cite{friedman_clock_2001} (as illustrated in Fig.~\ref{fig:virt_clk_tree}). This network can be generated in a recursive manner: The root clock buffer (stage 1) is assigned to the full set of available flip-flops. The second stage has two clock buffers which are driven by the buffer of the first stage. The full set of flip-flops is split in two disjoint sets with half the size and assigned to the two different clock buffers. For the next stage of clock buffers these sets are again divided in half and assigned to separated clock buffers which are driven by clock buffers of the previous stage. This process is repeated until the defined minimum fan-out of the clock buffer is reached. The example shown in Fig.~\ref{fig:virt_clk_tree} consists of a set of 9~flip-flops and a minimum fan-out of 2 which results in 3 levels of clock buffer. Due to the uneven number of flip-flops the actual fan-out of the clock buffer ranges from 2 to~3.

\subsection{Fault Injection Simulation Campaign}

With the previous described fault model a logic-level simulation based fault injection campaign can be performed. Therefore, the RTL model of the considered design and a testbench is needed. The testbench allows to verify the correct behaviour of the circuit. This can be done, for example, by monitoring and recording all outputs of the circuit. The record can be used as the golden reference and any difference is considered as a functional failure.
  
In the fault injection campaign faults are injected into the clock buffers of the clock network at a random point in time according to the described fault model. During each fault injection the changed and unchanged flip-flops are captured and stored. After the injection, the simulation is continued. The circuit output is monitored during the whole simulation and compared to the golden reference. If, according to the monitored output, no failure on the functional level was noted, the injected fault was masked and the correct function is verified. If the functional behaviour is different to the reference, the fault is considered as a functional failure. Thus, the Functional De-Rating factor of SETs in a clock buffer and the complete clock network can be computed. Further, by tracking the flip-flop changes which lead to a functional failure the vulnerability of the sequential logic can be calculated and thus, the most vulnerable flip-flops can be identified. This information can provide guidelines to the circuit designer to improve robustness of the clock distribution network. For example, techniques for selectively harden the most critical clock buffers are shown in~\cite{chellappa_90-nm_2012} and~\cite{mallajosyula_robust_2008}. Further, the \mbox{$\Delta$-TMR} technique can be used which hardens the sequential logic against SEUs, but also introduces delays into the data path in such a way the logic is protected against SETs in the clock signal~\cite{petrovic_design_2015}.

\section{Fault Injection Campaign}
\label{sec:results}

In this section the presented methodology and implemented fault model is shown on a practical example. Therefore, the circuit under test and the corresponding testbench is described. Afterwards, the functional failure rate for each clock buffer and for the complete network is computed. Additionally, the most vulnerable flip-flops related to SETs in the clock distribution network (CDN) are identified.

\subsection{Test Circuit, Testbench and Clock Distribution Network}

For this case-study the Ethernet 10GE~MAC Core from OpenCores is
used. The circuit implements the Media Access Control (MAC)
functions for 10\,Gbps operation as defined in the IEEE~802.3ae standard. The 10GE~MAC core has a 10\,Gbps interface
(XGMII TX/RX) to connect it to different types of Ethernet PHYs and
one packet interface to transmit and receive packets to/from the user
logic~\cite{andre_tanguay_10ge_2013}. The circuit consists of control
logic, state machines, FIFOs and memory interfaces. It is implemented
at the Register-Transfer Level (RTL) and is publicly available on
OpenCores.

The corresponding testbench writes several packets to the 10GE~MAC
transmit packet interface. As packet frames become available in the
transmit FIFO, the MAC calculates a CRC and sends them out to the
XGMII transmitter. The XGMII~TX interface is looped-back to the
XGMII~RX interface in the testbench. The frames are thus processed by
the MAC receive engine and stored in the receive FIFO. Eventually, the
testbench reads frames from the packet receive interface and prints
out the results~\cite{andre_tanguay_10ge_2013}. During the simulation
all sent and received packages to and from the core are monitored and 
recorded. This record is used as the golden reference for the fault 
injection campaign.

By performing a simple logic translation of the design, 1233\,flip-flops have been identified and matched with the corresponding RTL signal names. One virtual clock network was generated which groups flip-flops together according to their signal names and connects them to the same clock buffer. Additionally, 50 virtual clock networks were generated which connect the flip-flops to randomly selected clock buffers. The clock networks have a minimum fan-out of 16 flip-flops, which results in 7 stages and a total of 127 buffers with an actual fan-out from 19-20 flip-flops.

\subsection{Results for SETs in the Clock Distribution Network}

A fault injection campaign was performed to analyse the functional failure rate of SETs in the clock distribution network (CDN). Therefore, 170 SETs were injected in each of the 127 clock buffer of the different virtual clock networks. The faults were injected only during the active phase of the simulation, when packets are sent and received through the user packet interface. 

The overall results of the SET fault injection campaign are summarized in Table~\ref{tab:clk_set_res_assoc} and Table~\ref{tab:clk_set_res_rand}. Table~\ref{tab:clk_set_res_assoc} presents the results for the clock distribution network (CDN) which groups and connects flip-flops together based on their signal names. The number of reached, changed and unchanged flip-flops are listed for the entire campaign as well as the averaged number per injection. Further, the number of injections which lead to a functional failure is shown. Table~\ref{tab:clk_set_res_rand} presents the results for the same metrics but averaged over the 50 different random virtual clock networks. It was noted that the values for changed and unchanged values are identical. This can be explained by the fact that the pseudo random number generator always generates the same values to determine the injection time for each fault injection campaign. Thus, for each campaign the faults are injected at the same injection times and reaching the same flip-flops (via different buffers) which results in the same state changes. However, the functional failure rate is varying among the different random clock networks and especially in comparison to the not random clock network the functional failure rate differs by a factor of 2.

\begin{table}[htbp!]
    \caption{SET Fault Injection Campaign Results for CDN with Flip-Flops Grouped Together Based on the Signal Name}
    \label{tab:clk_set_res_assoc}
    \centering
    \begin{tabular}{lcc}
        \toprule
        & Total & Per Injection \\
        \midrule
        Injection Targets (Clock Buffers) & 127 & - \\
        Injected Faults (SET) & 21590 & - \\
        Reached FFs & 1467270 & 67.96 \\
        Changed FF States & 113008 & 5.23 \\
        Unchanged FF States  & 1354262 & 62.73 \\
        Functional Failure & 5423 & 25.12\,\%\\
        \bottomrule
    \end{tabular}
\end{table}

\begin{table}[htbp!]
    \caption{SET Fault Injection Campaign Results for CDN with Randomly Grouped Flip-Flops (Averaged over 50 CDNs)}
    \label{tab:clk_set_res_rand}
    \centering
    \begin{tabular}{lcc}
        \toprule
        & Total & Per Injection \\
        \midrule
        Injection Targets (Clock Buffers) & 127 & - \\
        Injected Faults (SET) & 21590 & - \\
        Reached FFs & 1467270 & 67.96 \\
        Changed FF States & 113008 & 5.23 \\
        Unchanged FF States  & 1354262 & 62.73 \\
        Functional Failure (averaged) & 11316 ($\pm$\,99) & 52.41\,\% ($\pm$\,0.46\,\%)\\
        \bottomrule
    \end{tabular}
\end{table}

\begin{figure}[htbp]
    \centering
    
    \begin{tikzpicture}
    \begin{axis}[
        xbar=0pt,
        /pgf/bar shift=0pt,
        ytick={0,...,60},
        axis y line*=none,
        axis x line*=bottom,
        tick label style={font=\tiny},
        width=.7125\linewidth,
        bar width=1mm,
        xmajorgrids,
        x label style={font=\footnotesize},
        xlabel={Functional Failure Rate},
        yticklabels={
            {\texttt{tx\_hold\_fifo0.fifo0.ctrl0.wr\_ptr(0)}},
            {\texttt{rx\_data\_fifo0.fifo0.ctrl0.rd\_ptr(0)}},
            {\texttt{tx\_hold\_fifo0.fifo0.ctrl0.rd\_ptr(0)}},
            {\texttt{rx\_data\_fifo0.fifo0.ctrl0.wr\_ptr(0)}},
            {\texttt{rx\_hold\_fifo0.fifo0.ctrl0.wr\_ptr(0)}},
            {\texttt{rx\_hold\_fifo0.fifo0.ctrl0.rd\_ptr(0)}},
            {\texttt{tx\_dq0.byte\_cnt(3)}},
            {\texttt{tx\_dq0.shift\_crc\_cnt(0)}},
            {\texttt{rx\_eq0.curr\_byte\_cnt(3)}},
            {\texttt{rx\_eq0.crc32\_d8(13)}},
            {\texttt{rx\_eq0.crc32\_d8(16)}},
            {\texttt{rx\_eq0.crc32\_d8(21)}},
            {\texttt{rx\_eq0.crc32\_d64(22)}},
            {\texttt{rx\_eq0.crc32\_d8(0)}},
            {\texttt{rx\_eq0.crc32\_d8(3)}},
            {\texttt{rx\_eq0.crc32\_d8(19)}},
            {\texttt{rx\_eq0.crc32\_d8(5)}},
            {\texttt{rx\_eq0.crc32\_d8(4)}},
            {\texttt{rx\_eq0.crc32\_d8(23)}},
            {\texttt{rx\_eq0.crc32\_d8(18)}},
            {\texttt{rx\_eq0.crc32\_d8(8)}},
            {\texttt{rx\_eq0.crc32\_d8(22)}},
            {\texttt{rx\_eq0.crc32\_d64(17)}},
            {\texttt{rx\_eq0.crc32\_d8(9)}},
            {\texttt{rx\_eq0.crc32\_d8(11)}},
            {\texttt{rx\_eq0.crc32\_d8(20)}},
            {\texttt{rx\_eq0.crc32\_d64(2)}},
            {\texttt{rx\_eq0.crc32\_d8(30)}},
            {\texttt{rx\_eq0.crc32\_d8(12)}},
            {\texttt{tx\_dq0.shift\_crc\_eop(0)}},
            {\texttt{rx\_eq0.crc32\_d64(3)}},
            {\texttt{rx\_eq0.crc32\_d8(17)}},
            {\texttt{rx\_eq0.crc32\_d8(7)}},
            {\texttt{rx\_eq0.crc32\_d8(26)}},
            {\texttt{rx\_eq0.crc32\_d64(16)}},
            {\texttt{rx\_eq0.crc32\_d64(21)}},
            {\texttt{tx\_dq0.curr\_state\_enc(0)}},
            {\texttt{rx\_eq0.crc32\_d64(13)}},
            {\texttt{rx\_eq0.crc32\_d8(6)}},
            {\texttt{rx\_eq0.crc32\_d8(24)}},
            {\texttt{tx\_dq0.crc32\_d64(20)}},
            {\texttt{tx\_dq0.crc32\_d64(22)}},
            {\texttt{rx\_eq0.crc32\_d8(2)}},
            {\texttt{tx\_dq0.crc32\_d64(26)}},
            {\texttt{tx\_dq0.crc32\_d64(3)}},
            {\texttt{rx\_eq0.crc32\_d8(28)}},
            {\texttt{rx\_eq0.crc32\_d64(28)}},
            {\texttt{tx\_dq0.crc32\_d64(5)}},
            {\texttt{rx\_eq0.crc32\_d64(19)}},
            {\texttt{rx\_eq0.crc32\_d8(27)}},
            {\texttt{tx\_dq0.crc32\_d64(4)}},
            {\texttt{rx\_eq0.crc32\_d64(0)}},
            {\texttt{tx\_dq0.crc32\_d64(18)}},
            {\texttt{rx\_eq0.crc32\_d8(15)}},
            {\texttt{rx\_eq0.crc32\_d64(24)}},
            {\texttt{tx\_dq0.crc32\_d64(21)}},
            {\texttt{rx\_eq0.xgxs\_rxd\_barrel(7)}},
            {\texttt{rx\_eq0.xgxs\_rxd\_barrel(4)}},
            {\texttt{tx\_dq0.crc32\_d64(24)}},
            {\texttt{rx\_eq0.crc32\_d8(14)}}
        },
        xmin=0,
        xmax=1,
        y=-1.75mm,
        enlarge y limits={abs=0.625},
        every axis plot/.append style={fill}
    ]
        \addplot[mybarcolor] coordinates {(0.42773109243697477,0)}; 
        \addplot[mybarcolor] coordinates {(0.4235294117647059,1)}; 
        \addplot[mybarcolor] coordinates {(0.4226890756302521,2)}; 
        \addplot[mybarcolor] coordinates {(0.42100840336134454,3)}; 
        \addplot[mybarcolor] coordinates {(0.42100840336134454,4)}; 
        \addplot[mybarcolor] coordinates {(0.41596638655462187,5)}; 
        \addplot[mybarcolor] coordinates {(0.41008403361344536,6)}; 
        \addplot[mybarcolor] coordinates {(0.3798319327731092,7)}; 
        \addplot[mybarcolor] coordinates {(0.3680672268907563,8)}; 
        \addplot[mybarcolor] coordinates {(0.29831932773109243,9)}; 
        \addplot[mybarcolor] coordinates {(0.29831932773109243,10)}; 
        \addplot[mybarcolor] coordinates {(0.29747899159663865,11)}; 
        \addplot[mybarcolor] coordinates {(0.2907563025210084,12)}; 
        \addplot[mybarcolor] coordinates {(0.28823529411764703,13)}; 
        \addplot[mybarcolor] coordinates {(0.28739495798319326,14)}; 
        \addplot[mybarcolor] coordinates {(0.28403361344537814,15)}; 
        \addplot[mybarcolor] coordinates {(0.2823529411764706,16)}; 
        \addplot[mybarcolor] coordinates {(0.280672268907563,17)}; 
        \addplot[mybarcolor] coordinates {(0.27647058823529413,18)}; 
        \addplot[mybarcolor] coordinates {(0.2739495798319328,19)}; 
        \addplot[mybarcolor] coordinates {(0.2739495798319328,20)}; 
        \addplot[mybarcolor] coordinates {(0.2714285714285714,21)}; 
        \addplot[mybarcolor] coordinates {(0.2689075630252101,22)}; 
        \addplot[mybarcolor] coordinates {(0.2689075630252101,23)}; 
        \addplot[mybarcolor] coordinates {(0.2689075630252101,24)}; 
        \addplot[mybarcolor] coordinates {(0.2672268907563025,25)}; 
        \addplot[mybarcolor] coordinates {(0.2630252100840336,26)}; 
        \addplot[mybarcolor] coordinates {(0.2571428571428571,27)}; 
        \addplot[mybarcolor] coordinates {(0.2571428571428571,28)}; 
        \addplot[mybarcolor] coordinates {(0.2571428571428571,29)}; 
        \addplot[mybarcolor] coordinates {(0.25630252100840334,30)}; 
        \addplot[mybarcolor] coordinates {(0.25210084033613445,31)}; 
        \addplot[mybarcolor] coordinates {(0.25210084033613445,32)}; 
        \addplot[mybarcolor] coordinates {(0.25126050420168067,33)}; 
        \addplot[mybarcolor] coordinates {(0.25126050420168067,34)}; 
        \addplot[mybarcolor] coordinates {(0.2504201680672269,35)}; 
        \addplot[mybarcolor] coordinates {(0.2495798319327731,36)}; 
        \addplot[mybarcolor] coordinates {(0.246218487394958,37)}; 
        \addplot[mybarcolor] coordinates {(0.246218487394958,38)}; 
        \addplot[mybarcolor] coordinates {(0.2453781512605042,39)}; 
        \addplot[mybarcolor] coordinates {(0.2445378151260504,40)}; 
        \addplot[mybarcolor] coordinates {(0.2445378151260504,41)}; 
        \addplot[mybarcolor] coordinates {(0.24369747899159663,42)}; 
        \addplot[mybarcolor] coordinates {(0.24201680672268908,43)}; 
        \addplot[mybarcolor] coordinates {(0.24201680672268908,44)}; 
        \addplot[mybarcolor] coordinates {(0.24033613445378152,45)}; 
        \addplot[mybarcolor] coordinates {(0.24033613445378152,46)}; 
        \addplot[mybarcolor] coordinates {(0.24033613445378152,47)}; 
        \addplot[mybarcolor] coordinates {(0.23949579831932774,48)}; 
        \addplot[mybarcolor] coordinates {(0.23865546218487396,49)}; 
        \addplot[mybarcolor] coordinates {(0.23865546218487396,50)}; 
        \addplot[mybarcolor] coordinates {(0.23781512605042016,51)}; 
        \addplot[mybarcolor] coordinates {(0.23445378151260504,52)}; 
        \addplot[mybarcolor] coordinates {(0.23361344537815126,53)}; 
        \addplot[mybarcolor] coordinates {(0.23361344537815126,54)}; 
        \addplot[mybarcolor] coordinates {(0.2319327731092437,55)}; 
        \addplot[mybarcolor] coordinates {(0.23109243697478993,56)}; 
        \addplot[mybarcolor] coordinates {(0.22941176470588234,57)}; 
        \addplot[mybarcolor] coordinates {(0.22941176470588234,58)}; 
        \addplot[mybarcolor] coordinates {(0.22857142857142856,59)};
    \end{axis}  
\end{tikzpicture}
    \vspace{-17.5pt}
    \caption{Most vulnerable flip-flops due to SETs in the CDN}
    \label{fig:clk_set_ff_fdr}
\end{figure}
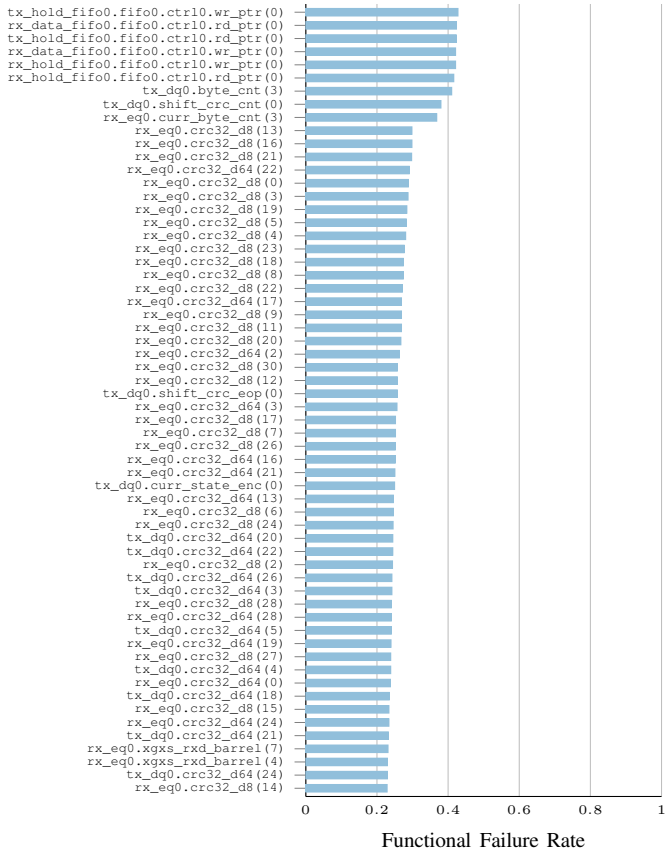

The most vulnerable flip-flops to SETs in the clock network are obtained by tracking the flip-flops which were reached and consequently changed their state due to an injected event and thus, led to a functional failure. Fig~\ref{fig:clk_set_ff_fdr} shows the most critical 5\,\% of the flip-flops for one of the randomly created clock networks, ranked by the individual functional failure rate. In case of selective mitigation these flip-flops should be considered for hardening with the $\Delta$-TMR technique~\cite{petrovic_design_2015}.

\subsection{Results for SEUs in the Sequential Logic}

The functional failure rate caused by SEUs in the flip-flops is obtained by a classical full flat statistical fault injection campaign. The SEU is emulated by modifying the stored value of a flip-flop at a random point in time during the simulation. Similar to the SET fault injection campaign, any difference in the send or received packages is considered as a functional failure of the application.

\begin{figure}[htbp]
    \centering
    
    \begin{tikzpicture}
    \begin{axis}[
        xbar=0pt,
        /pgf/bar shift=0pt,
        ytick={0,...,60},
        axis y line*=none,
        axis x line*=bottom,
        tick label style={font=\tiny},
        width=.7125\linewidth,
        bar width=1mm,
        xmajorgrids,
        x label style={font=\footnotesize},
        xlabel={Functional Failure Rate},
        yticklabels={
            {\texttt{rx\_hold\_fifo0.fifo0.ctrl0.wr\_ptr(3)}},
            {\texttt{rx\_data\_fifo0.fifo0.ctrl0.rd\_ptr(5)}},
            {\texttt{rx\_data\_fifo0.fifo0.ctrl0.rd\_ptr(7)}},
            {\texttt{rx\_data\_fifo0.fifo0.ctrl0.rd\_ptr(4)}},
            {\texttt{tx\_data\_fifo0.fifo0.ctrl0.rd\_ptr(4)}},
            {\texttt{rx\_hold\_fifo0.fifo0.ctrl0.rd\_ptr(4)}},
            {\texttt{rx\_data\_fifo0.fifo0.ctrl0.rd\_ptr(3)}},
            {\texttt{rx\_hold\_fifo0.fifo0.ctrl0.wr\_ptr(4)}},
            {\texttt{rx\_hold\_fifo0.fifo0.ctrl0.rd\_ptr(3)}},
            {\texttt{rx\_data\_fifo0.fifo0.ctrl0.wr\_ptr(3)}},
            {\texttt{rx\_data\_fifo0.fifo0.ctrl0.wr\_ptr(4)}},
            {\texttt{rx\_data\_fifo0.fifo0.ctrl0.wr\_ptr(5)}},
            {\texttt{rx\_data\_fifo0.fifo0.ctrl0.wr\_ptr(6)}},
            {\texttt{rx\_data\_fifo0.fifo0.ctrl0.wr\_ptr(7)}},
            {\texttt{rx\_data\_fifo0.fifo0.ctrl0.rd\_ptr(6)}},
            {\texttt{rx\_hold\_fifo0.fifo0.ctrl0.rd\_ptr(2)}},
            {\texttt{rx\_data\_fifo0.fifo0.ctrl0.rd\_ptr(1)}},
            {\texttt{rx\_data\_fifo0.fifo0.ctrl0.rd\_ptr(2)}},
            {\texttt{rx\_data\_fifo0.fifo0.ctrl0.rd\_ptr(0)}},
            {\texttt{rx\_hold\_fifo0.fifo0.ctrl0.rd\_ptr(1)}},
            {\texttt{rx\_hold\_fifo0.fifo0.ctrl0.rd\_ptr(0)}},
            {\texttt{tx\_data\_fifo0.fifo0.ctrl0.rd\_ptr(2)}},
            {\texttt{rx\_eq0.curr\_state(0)}},
            {\texttt{tx\_hold\_fifo0.fifo0.ctrl0.wr\_ptr(4)}},
            {\texttt{tx\_hold\_fifo0.fifo0.ctrl0.rd\_ptr(4)}},
            {\texttt{rx\_data\_fifo0.fifo0.ctrl0.wr\_ptr(0)}},
            {\texttt{rx\_data\_fifo0.fifo0.ctrl0.wr\_ptr(1)}},
            {\texttt{rx\_data\_fifo0.fifo0.ctrl0.wr\_ptr(2)}},
            {\texttt{tx\_hold\_fifo0.fifo0.ctrl0.rd\_ptr(3)}},
            {\texttt{rx\_hold\_fifo0.fifo0.ctrl0.wr\_ptr(0)}},
            {\texttt{rx\_hold\_fifo0.fifo0.ctrl0.wr\_ptr(2)}},
            {\texttt{rx\_hold\_fifo0.fifo0.ctrl0.wr\_ptr(1)}},
            {\texttt{wishbone\_if0.cpureg\_config0(0)}},
            {\texttt{tx\_hold\_fifo0.fifo0.ctrl0.rd\_ptr(2)}},
            {\texttt{tx\_hold\_fifo0.fifo0.ctrl0.rd\_ptr(1)}},
            {\texttt{tx\_data\_fifo0.fifo0.ctrl0.rd\_ptr(0)}},
            {\texttt{tx\_data\_fifo0.fifo0.ctrl0.rd\_ptr(1)}},
            {\texttt{tx\_dq0.curr\_state\_pad(0)}},
            {\texttt{tx\_data\_fifo0.fifo0.ctrl0.rd\_ptr(5)}},
            {\texttt{tx\_data\_fifo0.fifo0.ctrl0.rd\_ptr(3)}},
            {\texttt{tx\_data\_fifo0.fifo0.ctrl0.rd\_ptr(6)}},
            {\texttt{tx\_dq0.txhfifo\_wen}},
            {\texttt{tx\_dq0.byte\_cnt(6)}},
            {\texttt{tx\_hold\_fifo0.fifo0.ctrl0.wr\_ptr(3)}},
            {\texttt{tx\_dq0.crc32\_d8(9)}},
            {\texttt{tx\_dq0.crc32\_d8(25)}},
            {\texttt{tx\_dq0.crc32\_d8(28)}},
            {\texttt{tx\_hold\_fifo0.fifo0.ctrl0.wr\_ptr(2)}},
            {\texttt{tx\_dq0.crc32\_d8(16)}},
            {\texttt{tx\_dq0.crc32\_d8(29)}},
            {\texttt{tx\_dq0.crc32\_d8(24)}},
            {\texttt{tx\_dq0.crc32\_d8(26)}},
            {\texttt{tx\_dq0.crc32\_d8(7)}},
            {\texttt{tx\_dq0.crc32\_d8(6)}},
            {\texttt{tx\_dq0.crc32\_d8(17)}},
            {\texttt{tx\_dq0.crc32\_d8(3)}},
            {\texttt{tx\_dq0.crc32\_d8(5)}},
            {\texttt{tx\_dq0.crc32\_d8(4)}},
            {\texttt{tx\_dq0.crc32\_d8(8)}},
            {\texttt{tx\_dq0.crc32\_d8(27)}}
        },
        xmin=0,
        xmax=1,
        y=-1.75mm,
        enlarge y limits={abs=0.625},
        every axis plot/.append style={fill}
    ]
        \addplot[mybarcolor] coordinates {(1.0,0)}; 
        \addplot[mybarcolor] coordinates {(1.0,1)}; 
        \addplot[mybarcolor] coordinates {(1.0,2)}; 
        \addplot[mybarcolor] coordinates {(1.0,3)}; 
        \addplot[mybarcolor] coordinates {(1.0,4)}; 
        \addplot[mybarcolor] coordinates {(1.0,5)}; 
        \addplot[mybarcolor] coordinates {(1.0,6)}; 
        \addplot[mybarcolor] coordinates {(1.0,7)}; 
        \addplot[mybarcolor] coordinates {(1.0,8)}; 
        \addplot[mybarcolor] coordinates {(1.0,9)}; 
        \addplot[mybarcolor] coordinates {(1.0,10)}; 
        \addplot[mybarcolor] coordinates {(1.0,11)}; 
        \addplot[mybarcolor] coordinates {(1.0,12)}; 
        \addplot[mybarcolor] coordinates {(1.0,13)}; 
        \addplot[mybarcolor] coordinates {(1.0,14)}; 
        \addplot[mybarcolor] coordinates {(0.9941176470588236,15)}; 
        \addplot[mybarcolor] coordinates {(0.9941176470588236,16)}; 
        \addplot[mybarcolor] coordinates {(0.9941176470588236,17)}; 
        \addplot[mybarcolor] coordinates {(0.9941176470588236,18)}; 
        \addplot[mybarcolor] coordinates {(0.9941176470588236,19)}; 
        \addplot[mybarcolor] coordinates {(0.9941176470588236,20)}; 
        \addplot[mybarcolor] coordinates {(0.9529411764705882,21)}; 
        \addplot[mybarcolor] coordinates {(0.9470588235294117,22)}; 
        \addplot[mybarcolor] coordinates {(0.9,23)}; 
        \addplot[mybarcolor] coordinates {(0.9,24)}; 
        \addplot[mybarcolor] coordinates {(0.8823529411764706,25)}; 
        \addplot[mybarcolor] coordinates {(0.8823529411764706,26)}; 
        \addplot[mybarcolor] coordinates {(0.8823529411764706,27)}; 
        \addplot[mybarcolor] coordinates {(0.8705882352941177,28)}; 
        \addplot[mybarcolor] coordinates {(0.8235294117647058,29)}; 
        \addplot[mybarcolor] coordinates {(0.8235294117647058,30)}; 
        \addplot[mybarcolor] coordinates {(0.8235294117647058,31)}; 
        \addplot[mybarcolor] coordinates {(0.7352941176470589,32)}; 
        \addplot[mybarcolor] coordinates {(0.7176470588235294,33)}; 
        \addplot[mybarcolor] coordinates {(0.711764705882353,34)}; 
        \addplot[mybarcolor] coordinates {(0.7058823529411765,35)}; 
        \addplot[mybarcolor] coordinates {(0.7058823529411765,36)}; 
        \addplot[mybarcolor] coordinates {(0.6882352941176471,37)}; 
        \addplot[mybarcolor] coordinates {(0.6588235294117647,38)}; 
        \addplot[mybarcolor] coordinates {(0.6588235294117647,39)}; 
        \addplot[mybarcolor] coordinates {(0.6588235294117647,40)}; 
        \addplot[mybarcolor] coordinates {(0.6294117647058823,41)}; 
        \addplot[mybarcolor] coordinates {(0.6294117647058823,42)}; 
        \addplot[mybarcolor] coordinates {(0.5941176470588235,43)}; 
        \addplot[mybarcolor] coordinates {(0.5823529411764706,44)}; 
        \addplot[mybarcolor] coordinates {(0.5823529411764706,45)}; 
        \addplot[mybarcolor] coordinates {(0.5823529411764706,46)}; 
        \addplot[mybarcolor] coordinates {(0.5823529411764706,47)}; 
        \addplot[mybarcolor] coordinates {(0.5823529411764706,48)}; 
        \addplot[mybarcolor] coordinates {(0.5823529411764706,49)}; 
        \addplot[mybarcolor] coordinates {(0.5823529411764706,50)}; 
        \addplot[mybarcolor] coordinates {(0.5823529411764706,51)}; 
        \addplot[mybarcolor] coordinates {(0.5823529411764706,52)}; 
        \addplot[mybarcolor] coordinates {(0.5823529411764706,53)}; 
        \addplot[mybarcolor] coordinates {(0.5823529411764706,54)}; 
        \addplot[mybarcolor] coordinates {(0.5823529411764706,55)}; 
        \addplot[mybarcolor] coordinates {(0.5823529411764706,56)}; 
        \addplot[mybarcolor] coordinates {(0.5823529411764706,57)}; 
        \addplot[mybarcolor] coordinates {(0.5823529411764706,58)}; 
        \addplot[mybarcolor] coordinates {(0.5823529411764706,59)};
    \end{axis}  
\end{tikzpicture}
    \vspace{-17.5pt}
    \caption{Most vulnerable flip-flops due to SEUs in the sequential logic}
    \label{fig:seu_ff_fdr}
\end{figure}
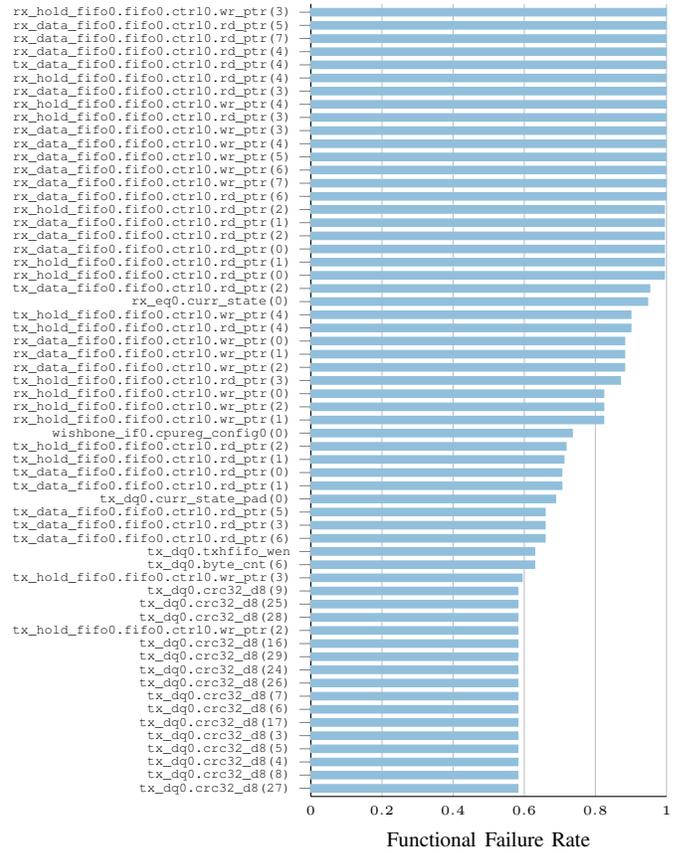

For the campaign 170 SEUs were injected in each of the 1233 flip-flops. 57150 of the injected faults showed a functional failure, which corresponds to an Functional De-Rating factor of 27.26\,\%. Table~\ref{tab:seu_ff_fdr}
summarizes the overall results of the SEU fault injection campaign. In Fig~\ref{fig:seu_ff_fdr} the most sensitive 5\,\% of the flip-flops are listed, ranked by the individual functional failure rate. This list is compared to the most vulnerable flip-flops due to SETs in the clock network in the next section.

\begin{table}[htbp!]
  \centering
  \caption{SEU Fault Injecton Campaign Results for the Sequential Logic}
  \label{tab:seu_ff_fdr}
  \begin{tabular}{lcc}
    \toprule
    & Total & Per Injection \\
    \midrule
    Injection Targets (FFs) & 1233 & - \\
    Injected Faults (SEU) & 209610 & - \\
    Functional Failure & 57150 & 27.26\,\%\\
    \bottomrule
  \end{tabular}
\end{table}

\subsection{Comparison and Discussion}

In order to compare the most vulnerable flip-flops related to SETs in the clock network, for each virtual clock network a list is created which contains the most critical 5\,\% of the flip-flops, ranked by the functional failure rate. Comparing these lists to each other, it has been noted that there is a big overlap of flip-flops. Each lists contains 60 flip-flops (the most critical 5\,\%) and the same 42 flip-flops can be found in each list, which results in an overlap of 70\,\%. In contrary, when comparing the most vulnerable flip-flops related to SETs in the clock network to the flip-flops related to SEUs almost no overlap can be found (3 flip-flops are common in each list which results in an overlap of 5\,\%). This is particularly important when selective hardening of the sequential logic is considered. For example, if there is a limited budget which can be used to harden flip-flops by using the $\Delta$-TMR technique, different flip-flops need to be taken into account in order to lower the functional failure rate related to both effects.

\bigskip

Furthermore, the functional failure rate due to SEUs in the sequential logic is compared to the error rate due to SETs in the clock network. Therefore, Table~\ref{tab:sum_fdr} summarizes the average Functional De-Rating factor per element. The Functional De-Rating factors are in the same order of magnitude but can be twice as much depending on the layout of the clock network. However, the number of sequential elements is 10 times higher and thus, the SEUs in flip-flops are the leading contributor to the overall functional failure rate of the circuit.

Considering further physical effects the Functional De-Rating factor can be combined with a FIT rate obtained from a characterized standard cell library. In~\cite{costenaro_practical_2013} FIT values for the NanGate FreePDK45 Open Cell Library~\cite{stine_freepdk_2007} were obtained by using dedicated tools and results from radiation testing. The average values for D-Flip-Flops and clock buffers show that the FIT value for the sequential logic is about 3 times higher, which further lowers the effect of SETs in the clock network.

Nonetheless, if a fully SEU hardened circuit is considered, the Functional De-Rating of the sequential logic is lower. Depending on the implementation of the hardened cells, the sensitivity is usually about one order of magnitude lower than the one of un-hardened cells. Taking this into account, the Functional De-Rating factor would is lowered by the same amount and thus, the functional failure rates getting closer to the failure rates due to SETs. This would mean that the SETs in the clock network are almost as significant as SEUs in the sequential logic.

\begin{table}[htbp!]
  \caption{Summary of the Functional Failure Rate Analysis \newline of the 10GE MAC circuit}
  \centering
  \label{tab:sum_fdr}
  \begin{tabular}{l@{}cccc}
    \toprule
    Element Type & \shortstack{Number \\ of Elements} & \shortstack{Average \\ FDR} & FIT & \shortstack{Functional \\ Failure \\ Rate} \\
    \midrule
    Flip-Flops           & 1233 & 0.27 & 161.75 & 53848 \\
    Clock Network        & 127  & 0.25 & 59.17  & 1878 \\
    Random Clock Network & 127  & 0.52 & 59.17  & 3907 \\
    \bottomrule
  \end{tabular}
\end{table}
\section{Conclusion}
\label{sec:conclusion}

This paper proposes a methodology to analyse how Single-Event Transients (SETs) in the clock distribution network are impacting the functional behaviour of a circuit. A methodology and a fault model were presented which implement the main radiation-induced effects in clock networks. The method enables the computation of the functional failure rate in a logic-level simulation based on the register-transfer level of the design. Thus, a faster evaluation can be performed than by simulating on the electrical level.

The approach was applied in a practical example. SETs were injected into the clock network of the circuit under test in a fault injection campaign. Thus, the functional failure rates of the clock network and the individual clock buffers were determined. Further, the most vulnerable flip-flops have been identified, which can be considered for selective mitigation techniques.

The proposed method uses a Virtual Clock Network which has the advantage that different clock network features can be evaluated with regard to Single-Event Effects in the clock network in an early design stage. However, the presented method can also be used in later design stages when the real clock network is available. This remains a topic for future work. In this paper two different types of clock network layouts were created. It has been noted that the layout can have a significant impact on the functional failure rate. Therefore, for future work further layouts and topologies of real clock distribution networks should be evaluated.

Finally, the functional failure rate due to SETs in the clock network has been compared to SEUs in the sequential logic. It was noted that there is almost no overlap looking at the most critical flip-flops. Further, the discussion has shown that the contribution of SETs in the clock network can be quite significant, if the circuit's sequential logic is only hardened against SEUs.

\bibliographystyle{IEEEtran}
\bibliography{IEEEabrv,bib/DTIS2019.bib}

\end{document}